\documentclass{appolb}
\usepackage{epsfig}
\usepackage{epsfig,cite,amsmath,amsfonts,amssymb}
\usepackage{graphicx}
\def\vev#1{\langle #1 \rangle}
\def\wino{\tilde{W}}
\def\higgsino{\tilde{H}}
\begin{document}
\eqsec  
\title{Non-minimal models of supersymmetric spontaneous CP violation
\thanks{Invited contribution to the proceedings of GUSTAVOFEST -
  Symposium in Honour of G. C. Branco ``CP Violation and the
  Flavour Puzzle''. Lisbon, 19-20 July 2005.}
}
\author{Ana M. Teixeira
\address{Departamento de F\'\i sica Te\'orica C-XI and
  Instituto de F\'\i sica Te\'orica C-XVI, \\
Universidad Aut\'onoma de
  Madrid, Cantoblanco, 28049 Madrid, Spain}
}
\maketitle
\begin{abstract}
We briefly review the question of spontaneous CP violation in some
models of weak interactions. The next-to-minimal supersymmetric
standard model, with two Higgs doublets and one gauge singlet, is one 
of the minimal extensions of the standard model where SCPV is viable. 
We analyse the possibility of spontaneous CP violation the 
next-to-minimal supersymmetric standard model with an extra
singlet tadpole term in the scalar potential, 
and confirm the existence of
phenomenologically consistent minima with non-trivial CP violating
phases. 
\end{abstract}
 
\section{Introduction}
The origin and realisation of the breaking of the CP symmetry in
nature remains an illusive topic in modern particle physics \cite{book} .
Although the standard model of particle physics (SM) succeeds in
accommodating the experimentally observed values of CP violation (CPV)
in meson physics though the Cabbibo-Kobayashi-Maskawa mechanism, many
questions remain to be answered. Among the most relevant issues are
the explanation of the observed baryon asymmetry of the Universe
(which requires new sources of CPV from physics beyond the SM), and
the strong CP problem. The strong CP problem, or equivalently, 
the smallness of the $\bar \theta$ parameter,
is the source of an important SM fine-tuning problem. In
fact, experimental bounds from the electric dipole moment (EDM) of the
electron, neutron and mercury atom force the flavour conserving phase
$\bar \theta$ to be as small as $10^{-10}$, 
a most unnatural value in the sense of 't Hooft \cite{Hooft}, 
since the SM Lagrangian does not acquire a new symmetry in
the limit where $\bar \theta \to 0$.

In the SM, CP violation is implemented in an explicit way: the
presence of complex Yukawa couplings breaks the CP symmetry at
Lagrangian level. 
However, this is not the only mechanism of CP
violation. A second and very attractive possibility lies in
spontaneous CP violation (SCPV), a framework where CP is originally a
symmetry of the Lagrangian, which is dynamically broken by complex
scalar vacuum expectation values (VEVs).

The scenario of SCPV (or soft CPV), 
originally proposed by T. D. Lee \cite{tdlee},
is based on the follow principles: originally, the Lagrangian
describing the $SU(2)_L \times U(1)_Y$ theory is invariant under a
transformation that one can associate with the CP symmetry. During the
process of electroweak (EW) symmetry breaking, the scalar field that
is responsible for the breaking $SU(2)_L \times U(1)_Y \to
U(1)_{\mathrm{em}}$ acquires a complex VEV, so that after EW symmetry
breaking the Lagrangian is not longer CP invariant. The effects of
SCPV are finite in renormalisable theories \cite{tdlee}.

The motivations for SCPV are extensive, and the strongest are related
with providing a solution to the strong CP problem, as well as 
establishing a
connection between the breaking of the CP symmetry at very high
energies (where it could be understood in the framework of some
fundamental theory of particle physics - string theory, for instance)
and low energy phenomenology.
Regarding the strong CP problem, if CP is imposed as a symmetry of the
Lagrangian prior to spontaneous EW symmetry breaking, at the tree-level
one directly obtains $\bar \theta=0$. 
Within the context of supersymmetric (SUSY) theories, SCPV emerges as
a natural candidate to explain the SUSY CP-problem. SUSY models as the
minimal supersymmetric standard model (MSSM) introduce a vast array of
new CP violating phases, some of them flavour-conserving (as those
associated with the $\mu$-term and the soft gaugino masses). These
phases also generate sizable contributions to the EDMs, and are forced
to be very small to ensure compatibility with experiment, leading
to another naturalness problem - the SUSY CP-problem. 
Under SCPV, all these phases are set to zero at some intermediate
scale via the imposition of a symmetry, which is then softly broken, thus
allowing to evade 't Hooft's criteria \cite{bbsegre}.   
Further motivation for SCPV stems from string theory, where it has
been shown that in string perturbation theory CP exists as a symmetry
that could be spontaneously broken. In this framework, CP violation
could be induced by non-trivial properties of the manifold, or by CP
non-invariant compactification boundary conditions. Alternatively, CPV
could also originate from complex VEVs of the moduli fields. It has
also been argued that CP can be a gauge symmetry 
in string theory \cite{stringCP},
which is spontaneously broken at a high scale, the effects then being
fed to the Yukawa couplings in the superpotential, and to the soft
breaking terms.

Effective low energy models with a minimal Higgs field content, as is
the case of the SM, are not viable scenarios for SCPV. In fact, SCPV
requires at least two Higgs doublets, as is the case of the Lee model. 
Supersymmetric models emerge as interesting candidate scenarios for
SCPV. Not only one has at least a minimal content of two Higgs
doublets, but as discussed before, SCPV also solves issues
as the SUSY CP-problem. In section \ref{sect:2} we shall present a
short overview of SCPV in SUSY scenarios. In section \ref{sect:3},
we focus our discussion in the specific case of the next-to-minimal
supersymmetric standard model with a $\mathbb{Z}_3$ symmetry in the
superpotential \cite{Lisbon}, and an additional tadpole term in the
scalar potential. 
After defining the model, we address
the minimisation of the scalar potential, and study the mass spectrum.
Finally, in section \ref{sect:4}, we consider the contributions to
indirect CP violation in the neutral Kaon sector (parametrised by
$\varepsilon_K$), and present our numerical results. A summarising
outlook is presented in the Conclusions.

\section{Models of SUSY spontaneous CP violation}\label{sect:2}

Even though the original Lee model appeared to encompass all the
necessary ingredients for minimally extending the SM in order to
arrive at an electroweak model that softly broke CP, many were the
phenomenological problems that plagued it. Among them were excessive
contributions to leptonic and nucleon EDMs, and most important, the
existence of flavour-changing neutral Yukawa interactions
(FCNYI). FCNYI are a usual feature of multi-Higgs doublet models, and
typically induce excessive (and experimentally incompatible)
contributions to neutral meson mixing and rare decays. The
most elegant way to suppress them is to assume the existence of some
underlying mechanism (a symmetry) that removes the dangerous Higgs-quark
couplings. As shown in \cite{nfc}, the only way to have natural
flavour conservation is by imposing that each of the scalar doublets
only couples to a quark of a given charge. This can be achieved by
imposing a number of discrete symmetries on the Lagrangian. Albeit,
once such symmetries are imposed, SCPV in no longer possible.

One can further increase the number of the Higgs doublets: in the
Branco model \cite{3hdm}, which contains three doublets and has natural flavour
conservation, CP can be spontaneously broken and scalar particle
interactions are the only source of CP violation. One should also
mention that SCPV can also occur in models where, instead of just
enlarging the Higgs sector, one has additional fermions or an extended
gauge sector (e.g. extended left-right symmetric models, models with
additional heavy exotic fermions, vector-like quark models - see
\cite{book} for a comprehensive review).

The MSSM emerges as an appealing scenario for SCPV, since it has by
construction two Higgs doublets, and natural flavour conservation is
automatic. Nevertheless, it is well known that, at the tree-level,
SCPV does not occur in the MSSM \cite{HHG}. On the other hand, radiative
corrections 
can generate CP violating operators \cite{Maekawa} but then, according to the
Georgi-Pais theorem on radiatively broken global symmetries \cite{GP}, one
expects to have light states in the Higgs spectrum \cite{Pomarol}, which are
excluded by LEP \cite{LEP,LEPHA,LEPCH}.
Thus, it is of interest to consider simple extensions of the MSSM,
such as a model with at least one gauge singlet field ($N$) in
addition to the two Higgs doublets - the so-called 
the next-to-minimal supersymmetric standard model (NMSSM), and
investigate whether SCPV can be achieved in this class of models.

The NMSSM \cite{gen:nmssm} 
is a particularly appealing SUSY model, since it allows to
solve another MSSM naturalness problems, the so-called
$\mu$ problem \cite{muprob}. The $\mu$ problem arises from the presence
of a mass term for the Higgs fields in the superpotential, $\mu H_1
H_2$. The only natural values for the $\mu$ parameter are either zero
or the Planck scale. The first is experimentally excluded, since it
leads to an unacceptable axion once the EW symmetry is broken. The
second is equally unpleasant, since it reintroduces the hierarchy
problem. Although there are several explanations for an
$\mathcal{O}(M_W)$ value for the $\mu$ term, all are in extended
frameworks. 
The NMSSM offers a simple yet elegant
solution via the presence of a trilinear dimensionless coupling in
the superpotential, $\lambda N H_1 H_2$. When the scalar component of
$N$ acquires a VEV of the order of the SUSY breaking scale, 
an effective $\mu$ term is dynamically generated.
This realisation of the NMSSM, where the superpotential is
invariant under a $\mathbb{Z}_3$ symmetry is the simplest SUSY extension of the
SM where the EW scale exclusively originates from SUSY breaking.

From the point of view of SCPV, and in spite of all its many attractive
features, the NMSSM presents some problems. It has been shown that in
the simplest $\mathbb{Z}_3$ invariant version, there is no SCPV at the
tree-level. Even though CP violating extrema can be found, these are
maxima, and not minima of the potential, associated with tachyonic
Higgs states (no-go theorem) \cite{Jorge}. 
Regarding the possiblity of radiatively induced SCPV in the $\mathbb{Z}_3$
symmetric NMSSM, the situation is very similar to the MSSM case, since
the CPV minima are associated to very light Higgs states, which are
difficult to accomodate with experimental data \cite{BB, light:babuhaba}.

If one abandons the prospect of solving the $\mu$ problem, and allows
the presence of dimensionfull, SUSY conserving terms in the
superpotential, SCPV is indeed viable. As shown 
in \cite{Pomarol2, Lisbon}, one
can find CP violating minima of the potential which are in agreement
with experimental data on the Higgs sector, and can successfully
account for the observed value of $\varepsilon_K$. 
However, these models encompass an important theoretical drawback,
since in the absence of a global symmetry under which the singlet
field is charged, divergent singlet tadpoles proportional to
$M_{\mathrm{Planck}}$, generated by non-renormalizable higher order
interactions, can appear in the effective scalar potential
\cite{nonrenor}. These would lead to a destabilisation of the
hierarchy between EW the Planck scale. On the other hand, imposing
a discrete symmetry to overcome the latter problem leads to disastrous
cosmological domain walls 

Still, there is a possible solution to this controversial puzzle,
which consists in finding NMSSM models that while $\mathbb{Z}_3$-violating,
have a $\mathbb{Z}_3$ conserving superpotential. As pointed out in
\cite{Tamvakis:MNSSM}, using global discrete
$R$-symmetries for the complete theory - including non-renormalisable
interactions - one could construct a $\mathbb{Z}_3$ invariant renormalisable
superpotential and generate a $\mathbb{Z}_3$ breaking non-divergent singlet
tadpole term in the scalar potential.
In addition to being free of both stability and domain wall problems,
these models present a rather unique feature: they are a viable
scenario for SCPV, where one can obtain the observed value of 
$\varepsilon_K$, and have at the same time compatibility with
experimental data \cite{rcsp}. 
In the following section we proceed to analyse this
class of models in greater detail.

\section{Spontaneous CP violation in the NMSSM}\label{sect:3}
In this section, we will address the possibility of SCPV in the NMSSM
with an extra singlet tadpole term in the effective potential, taking
into account the constraints on Higgs and sparticle masses.
We begin by briefly describing the model, focusing of the Higgs scalar
potential and its minimisation. We then proceed to compute the masses
of the Higgs states, including radiative corrections, and present a
short numerical analysis of the neutral Higgs spectrum.

\subsection{The scalar Higgs potential}
We consider the most general form of the superpotential where, in
addition to the Yukawa couplings for quark and leptons (as in the
MSSM), we have the following Higgs couplings
\begin{equation} \label{supot} 
W_{\mathrm{Higgs}} = \lambda \hat{H}_1 \hat{H}_2 \hat{N} + 
\frac{1}{3} \kappa \hat{N}^3 \; ,
\end{equation}
where $\hat N$ is a singlet superfield, and $H_{1,2}$ are the usual
MSSM HIggs doublets. After EW symmetry breaking, 
the scalar component of $\hat{N}$ acquires
a VEV, $x = |\vev N|$, thus generating an effective $\mu$ term
\begin{equation} \label{defmu}
\mu \equiv \lambda x \; .
\end{equation}
As mentioned in the previous section, a possible means to overcome the
domain wall problem without spoiling the quantum stability of the
model is by replacing the $\mathbb{Z}_3$ symmetry by a set of discrete 
$R$-symmetries, broken by the soft SUSY breaking
terms \cite{Tamvakis:MNSSM}. At low energy, the additional non-renormalisable
terms allowed by the $R$-symmetries generate an extra linear term for the
singlet in the effective potential, through tadpole loop diagrams
\begin{equation} \label{Vtad}
V_{\mathrm{tadpole}} = - \xi^3 N + \mathrm{H.c.} \; ,
\end{equation}
where $\xi$ is of the order of the soft SUSY breaking terms ($\lesssim$ 1
TeV)\footnote{Since our approach is phenomenological, 
we therefore take $\xi$ as a free
parameter, without considering the details of the non-renormalisable
interactions generating it.}.
In addition to $V_{\mathrm{tadpole}}$ the tree-level Higgs potential
comprises the usual $D-$ and $F$-terms, as well as soft-SUSY breaking
interactions. The latter are given by
\begin{equation}\label{treepot}
V_\mathrm{soft} = m_{H_i}^2 |H_i|^2 + m_N^2 |N|^2 +
\left( \lambda A_\lambda N H_1 H_2 + 
\frac{1}{3} \kappa A_\kappa N^3 + \mathrm{H.c.}
\right). 
\end{equation}
In the above, we take the soft-SUSY breaking terms $m_{H_1}, m_{H_2},
m_N, A_\lambda, A_\kappa$ as free parameters at the weak scale.
We also assume that the
Lagrangian is CP invariant, which means that all the parameters appearing in
Eqs.~(\ref{Vtad},\ref{treepot}) are real. 

After spontaneous EW symmetry breaking, the neutral Higgses acquire 
complex VEVs that spontaneously break CP: 
\begin{equation} 
\vev{H_1^0} = v_1 e^{i\varphi_1} \; , \quad \vev{H_2^0} = v_2 e^{i\varphi_2} 
\; , \quad
\vev{N} = x e^{i\varphi_3} \; ,
\end{equation}
where $v_1, v_2, x$ are positive and $\varphi_1, \varphi_2, \varphi_3$
are CP violating phases. However, only two of these phases are physical. They
can be chosen as
\begin{equation} 
\theta = \varphi_1+\varphi_2+\varphi_3 \quad \mathrm{and} 
\quad \delta = 3 \varphi_3 \; .
\end{equation}

\subsection{CP violating minima of the scalar potential}
From the tree-level scalar potential of Eqs.\ref{Vtad},\ref{treepot}),
together with the associated $D$- and $F$- terms (see \cite{rcsp}), one
can derive the five minimisation equations for the VEVs and phases 
$v_1, v_2, x, \theta, \delta$. 
These can be used to express the soft parameters $m_{H_1}, m_{H_2}$,
$m_N, A_\lambda, A_\kappa$ in terms of $v_1, v_2, x, \theta, \delta$:

\begin{eqnarray} \label{mineqtree}
\frac{\partial V_\mathrm{tree}}{\partial v_1} = 0 \quad \Rightarrow \quad
m_{H_1}^2 & = & -\lambda^2 \left( x^2 + v^2\sin^2\beta \right) - 
\frac{1}{2} M_Z^2 \cos 2\beta \nonumber \\
& & - \lambda x \tan\beta \left( \kappa x \cos(\theta-\delta) + 
A_\lambda \cos\theta \right) \; , \nonumber \\
\frac{\partial V_\mathrm{tree}}{\partial v_2} = 0 \quad \Rightarrow \quad
m_{H_2}^2 & = & -\lambda^2 \left( x^2 + v^2\cos^2\beta \right) + 
\frac{1}{2} M_Z^2 \cos 2\beta \nonumber \\
& & - \lambda x \cot\beta \left( \kappa x \cos(\theta-\delta) + 
A_\lambda \cos\theta \right) \; , \nonumber \\
\frac{\partial V_\mathrm{tree}}{\partial x} = 0 \quad \Rightarrow \quad
m_{N}^2 & = & -\lambda^2 v^2 - 2\kappa^2 x^2 - \lambda \kappa v^2 \sin
2\beta \cos(\theta-\delta) \nonumber \\
& & - \frac{\lambda A_\lambda v^2}{2x} \sin 2\beta \cos\theta - 
\kappa A_\kappa x \cos\delta +
\frac{\xi^3}{x} \cos(\delta/3) \; , \nonumber \\
\frac{\partial V_\mathrm{tree}}{\partial \theta} = 0 \quad \Rightarrow \quad
A_\lambda & = & - \frac{\kappa x \sin(\theta-\delta)}{\sin\theta} 
\; , \nonumber \\
\frac{\partial V_\mathrm{tree}}{\partial \delta} = 0 \quad \Rightarrow \quad
A_\kappa & = & \frac{3 \lambda\kappa x v^2 \sin 2\beta
  \sin(\theta-\delta) + 2 \xi^3 \sin(\delta/3)}{2 \kappa
x^2 \sin\delta} \; ,
\end{eqnarray}
with $\tan\beta = v_2 / v_1$, $v = \sqrt{v_1^2+v_2^2} = 174$ GeV and $M_Z$
the $Z$ boson mass. The above relations allow us to use 
$\tan\beta, x, \theta$ and $\delta$ 
instead of $m_{H_1}, m_{H_2}, m_N, A_\lambda, A_\kappa$ as free parameters.
Once EW symmetry is spontaneously broken, we are left with five neutral Higgses
and a pair of charged Higgses. The neutral Higgs fields can be rewritten in
terms of CP eigenstates
\begin{align}
&H_1^0 =  e^{i \varphi_1} \left \{ v_1 + \frac{1}{\sqrt{2}} 
(S_1 + i \sin\beta P)
\right \}, \ 
H_2^0 =  e^{i \varphi_2} \left \{ v_2 + \frac{1}{\sqrt{2}} 
(S_2 + i \cos\beta P)
\right \} ,\nonumber \\
&\quad \quad \quad
N  =  e^{i \varphi_3} \left \{ x + \frac{1}{\sqrt{2}} 
(X + i Y) \right \} \; , 
\end{align}
where $S_1, S_2, X$ are the CP-even components, $P, Y$ are the CP-odd
components, and we have already rotated away the CP-odd would-be
Goldstone boson. The detailed expression for the mass matrix of the
scalar and pseudoscalar states can be found in \cite{rcsp}.
Here, it suffices to stress that from inspection of the neutral Higgs
boson mass matrix one can conclude that there is no CP violation in
the Higgs doublet sector, while for $\theta \neq \delta$ CP violating
mixings between singlet and doublets can appear. Moreover, the
presence of terms proportional to $\xi^3$ in the diagonal singlet entries
have the effect of lifting potentially negative eigenvalues, thus
allowing to evade the no-go theorem \cite{Jorge}.
Thus, SCPV is possible already at tree level for $\xi \neq 0$.

\subsection{Mass spectrum}
Although we will not enter in a detailed analysis here, let us mention
that, as occurs in the MSSM, radiative corrections to the Higgs masses
are very important, and play a crucial role in the SCPV
mechanism, as they generate CP violating operators \cite{Maekawa, BB}.
In the analysis of \cite{rcsp}, we have taken into account one-loop
contributions \cite{loop1} associated to top-stop and bottom-sbottom
loops\footnote{We took into account identical corrections when computing
  the charged Higgs boson mass.}. It is
relevant to refer that the one-loop terms give contributions to 
the minimisation conditions of Eq.~(\ref{mineqtree}), which should be
changed in order to incorporate the corrections. Regarding two-loop
corrections, 
we considered the dominant terms \cite{loop2} which are proportional to 
$\alpha_s h_t^4$ and $h_t^6$, taking
only the leading logarithms into account \cite{CH}. 
Once all these contributions are taken into account, one obtains a rather
complicated $5 \times 5$ mass matrix for the neutral Higgs fields, which can
only be numerically diagonalised \cite{rcsp}.

The next step consists in investigating whether it is possible to have 
SCPV in the NMSSM with the extra tadpole term for the singlet, 
given the exclusion limits on the
Higgs spectrum from LEP \cite{LEP,LEPHA,LEPCH}. In order to do so, we perform a
numerical scanning of the parameter space of the model.
Using the minima equations to replace the soft SUSY
breaking terms by the Higgs VEVs and phases, and using the effective
$\mu$ term as a free parameter instead of the singlet VEV, the free
parameters of the tree level Higgs mass matrix are now given by
\begin{equation} \label{freepar}
\lambda \; , \; \kappa \; , \; \tan\beta \; , \; \mu \; , \; \xi \; , 
\; \theta \; , \; \delta \; .
\end{equation}
Requiring the absence of Landau poles for $\lambda$ and $\kappa$
below the GUT scale translates
into bounds for the low-energy values of the couplings $\lambda$ and
$\kappa$. For $\tan\beta \leq 10$ and $m_t^{\rm
pole} = 175$ GeV, one finds $\lambda_{\rm max} \sim 0.65$ and 
$\kappa_{\rm max} \sim 0.6$. 
This also yields a lower bound for $\tan\beta$, namely $\tan\beta \gtrsim 2.2$.

Taking $M_{\rm SUSY} = 350$ GeV and
assuming a maximal mixing scenario for the stops, 
we have a numerical scanning on the free parameters, which were
randomly chosen in the following intervals:
\begin{align} \label{param}
& 
0.01 < \lambda < 0.65 \,,\quad 0.01 < \kappa < 0.6\,, \quad
0 < \xi < 1\,,\quad 100\, \mathrm{GeV} < \mu < 500\, \mathrm{GeV}\,,
\nonumber \\
& \quad \quad \quad \quad 
2.2 < \tan\beta < 10\,, \quad -\pi < \theta < \pi\,, \quad
-\pi < \delta < \pi\,.
\end{align}
For each point we computed the neutral 
Higgs masses and couplings, as well as the
charged Higgs, stop and the chargino masses, applying all the
available experimental constraints on these particles from LEP
\cite{LEP,LEPHA,LEPCH,LEPCHARGI,LEPSTOP}.
First of all, one verifies that in the $\mathbb{Z}_3$ invariant limit, where 
$\xi=0$, it is extremely difficult, if not impossible, to satisfy the
LEP constraints on the Higgs sector with non-zero CPV phases.
This is easily understood from the fact that in the limit where $\xi$
goes to zero, SCPV is no longer possible at tree level \cite{Jorge}, and
although viable when radiative corrections are included, the Georgi-Pais
theorem \cite{GP} predicts the appearance of light states in the Higgs
spectrum, already excluded by LEP. 
On the other hand, when $\xi \neq 0$, CP can be spontaneously broken
already at the tree level. Taking radiative corrections up to the
dominant two-loop terms, we obtained that large portions of the
parameter space of Eq.~(\ref{param}) complied with all the imposed
constraints for any values of the CP violating phases $\theta$ and
$\delta$. In other words, it is possible to have SCPV in the NMSSM
already at the tree level, as noted in the previous section.
\begin{figure}
\begin{center}
\vspace*{4mm}
\includegraphics[height=45mm]{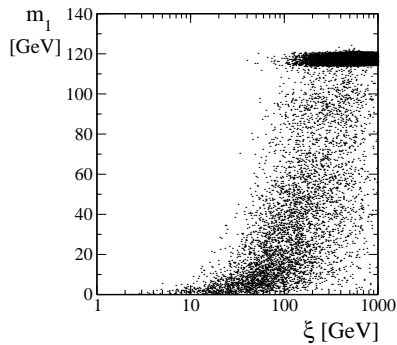}
\caption{Mass of the lightest Higgs as a function of $\xi$, for a  $m_t^{\rm
pole} = 175$ GeV, $M_{\rm SUSY} = 350$ GeV and maximal stop mixing. The other
parameters are randomly chosen as in Eq.~(\ref{param}).} \label{fig1}
\end{center}
\end{figure}

In Fig.~\ref{fig1} we display the mass of the 
lightest Higgs, $m_1$, as a function
of the tadpole parameter $\xi$, with the other parameters randomly chosen as in
Eq.~(\ref{param}). 
We can see that
small values of $\xi$ are associated with a very light mass for the $h_1$
state. Such light Higgs states are not excluded by current experimental bounds
since their reduced coupling to the SM gauge bosons is small enough to
avoid detection.

Before concluding this section, let us comment that regarding the LEP
bounds considered, in addition to bounds on the $Zh_i h_j$ reduced coupling 
\cite{LEPHA}, we have also
taken into account the LEP limit on the charged Higgs mass
\cite{LEPCH}.
As we will see in the next section, charginos play an important role in the
computation of $\varepsilon_K$. 
The tree level chargino mass matrix in the $(\wino,\higgsino)$ basis reads
\begin{equation}
{\cal M}_{\tilde \chi^{\pm}} = \left( 
\begin{array}{cc} 
M_2 & \sqrt{2} M_W \sin\beta e^{-i\theta} \\
\sqrt{2} M_W \cos\beta & - \lambda x 
\end{array} \right) \; ,
\end{equation}
where $M_2$ is the soft wino mass, which was randomly scanned in the
interval  
\begin{equation} \label{M2range}
100 \,\mathrm{GeV} < M_2 <250 \,\mathrm{GeV}\,.
\end{equation}
We also applied the LEP bound on the chargino and stop masses
\cite{LEPCHARGI, LEPSTOP}.

\section{$\varepsilon_K$ in the NMSSM}\label{sect:4}
In the framework of the NMSSM with SCPV, all the SUSY parameters are
real. Furthermore the SM does not provide any contribution to 
any of the CP violation observables, since the CKM matrix is real (and
the unitarity trinagle is thus flat). Even
so, the physical phases of the Higgs doublets and singlet appear in the scalar
fermion, chargino and neutralino mass matrices, as well as in several
interaction vertices. In what follows our aim is to investigate 
whether or not these physical phases can account for the experimental 
value of $\varepsilon_K=(2.284 \pm 0.014) \times 10^{-3}$~\cite{pdg2004}.

\subsection{Dominant contributions to $\varepsilon_K$}
\begin{figure}
\begin{center}
\includegraphics[height=7cm]{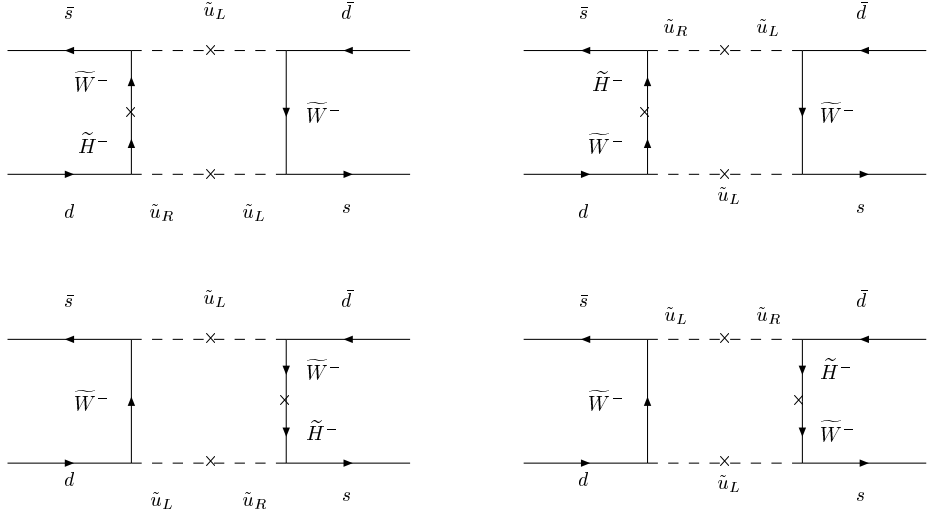}
\caption{Box diagrams associated with the leading chargino contribution to
$\varepsilon_K$.} \label{fig:eK:4box}
\end{center}
\end{figure}

Let us now proceed to compute the contributions to the indirect CP violation
parameter of the kaon sector, namely $\varepsilon_K$, which is defined as
\begin{equation} \label{eK:eq:def}
\varepsilon_K\simeq \frac{e^{i\pi/4}}{\sqrt{2}} \frac{\operatorname{Im}
\mathcal{M}_{12}}{\Delta m_K} \; .
\end{equation}
In the latter $\Delta m_K$ is the long- and short-lived kaon mass
difference, and $\mathcal{M}_{12}$ is the off-diagonal element of the neutral
kaon mass matrix, related to the effective hamiltonian that governs $\Delta
S=2$ transitions as
\begin{equation} \label{eK:M12:def}
\mathcal{M}_{12} = \frac{\langle K^0|\mathcal{H}_{\text{eff}}^{\Delta S=2}
|\bar{K}^0 \rangle}{2m_K} \; , \quad \text{with} \qquad
\mathcal{H}_{\text{eff}}^{\Delta S=2} = \sum_i c_i {\mathcal O}_i \; .
\end{equation}
In the above $c_i$ are the Wilson coefficients and ${\mathcal O}_i$ the local
operators. In the presence of SUSY contributions, the Wilson coefficients can
be decomposed as $c_i= c_i^W + c_i^{H^\pm}+ c_i^{\tilde{\chi}^\pm} +
c_i^{\tilde{g}} + c_i^{\tilde{\chi}^0}$. As discussed in Ref.~\cite{Lisbon}, in
the present class of models where there are no contributions from the SM, the
chargino mediated box diagrams give the leading supersymmetric contribution,
and the $\Delta S=2$ transition is largely dominated by the $(V-A)$ four
fermion operator $\mathcal{O}_1$. The contributions to $\varepsilon_K$
become more transparent if one works in the weak basis for the
$\tilde{W}-\tilde{H}$. It can be seen that the 
leading contribution arises from the box diagrams depicted
in Fig.~\ref{fig:eK:4box}. 
In the limit of degenerate masses for the left-handed
up-squarks,  $\operatorname{Im} \mathcal{M}_{12}$ is given by~\cite{Lisbon}
\begin{eqnarray} \label{eK:M12:equation}
\operatorname{Im} \mathcal{M}_{12}  & = & \frac{2 G_F^2 f^2_K m_K m_W^4
}{3\pi^2\vev{m_{\tilde{q}}}^8} (V_{td}^*V_{ts}) m_t^2 \left| e^{i\theta} \;
m_{\tilde{W}} - \cot\beta \; m_{\tilde{H}} \right| \nonumber \\
& & \times \left\{ \Delta A_U \sin[\varphi_{\chi}-\theta] \;
{(M^2_{\tilde{Q}})}_{12} \; I(r_{\tilde{W}}, r_{\tilde{H}}, r_{\tilde{u}_L},
r_{\tilde{t}_R}) \right\} \; ,
\end{eqnarray}
where $f_K$ is the Kaon decay constant and $m_K$ the Kaon
mass~\cite{pdg2004}; $V_{ij}$ are the $V_{\mathrm{CKM}}$ 
elements, whose numerical values
($V_{td}=0.0066$ and $V_{ts}=-0.04$) reflect the fact that we are dealing with
a flat unitarity triangle; $\vev{m_{\tilde{q}}}$ is the average squark mass,
which we take equal to $M_{\rm SUSY}$; $m_{\tilde{W}} = M_2$ is the wino mass,
$m_{\tilde{H}} = \mu$ is the higgsino mass and $\varphi_{\chi} = \arg (
e^{i\theta}m_{{\wino}} - \cot\beta\; m_{{\higgsino}} )$. 
${(M^2_{\tilde{Q}})}_{12}$ parametrises the non-universality in the
$LL$ soft breaking masses\footnote{The values for
  ${(M^2_{\tilde{Q}})}_{12}$ were taken in agreement with the bounds
  from~\cite{MQ12}}. The difference 
$\Delta A_U\equiv A_U^{13}-A_U^{23}$ reflects 
the non-universality in the soft trilinear terms, which is crucial 
in order to succeed in complying with the
experimental data. Finally, $I$ is the loop function, with
$r_i=m_i^2/\vev{m_{\tilde{q}}}^2$~\cite{Lisbon}.

\subsection{Numerical results and discussion}
As shown in \cite{rcsp}, a thorough scan of the parameter space confirms
that it is indeed possible to satisfy the minimisation conditions of the Higgs
potential, have an associated Higgs spectrum compatible with LEP searches and
still succeed in generating the observed value of $\varepsilon_K$. 
In the numerical analysis, the free parameters of the model were taken
as in Eqs.(\ref{param},\ref{M2range}), with $m_t^{\rm pole} = 175$ GeV and
maximal stop mixing, as in the previously discussed. Moreover, we have
taken $100\, \mathrm{GeV} \lesssim M_2 \lesssim 250$ GeV, and $M_{\rm
  SUSY}=350$ GeV (a value that reflects the compromise between the need
to generate a sufficiently heavy Higgs spectrum and at the same time 
account for the observed $\varepsilon_K$).
As noted in \cite{rcsp}, 
saturating the observed value of $\varepsilon_K$ favours a regime of
low $\tan \beta$, with the maximal values of $\varepsilon_K$ being
obtained for $\tan \beta \lesssim 3.8$.

\begin{figure}
\begin{center}
\includegraphics[height=45mm]{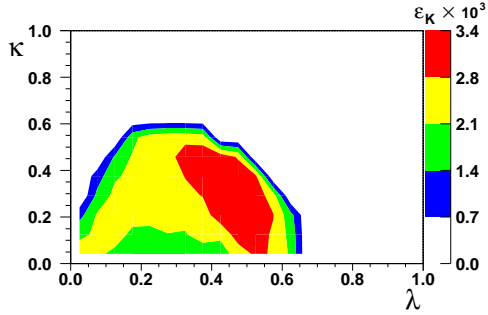}
\caption{Contours for the maximum value of $\varepsilon_K$ 
in the $\lambda$--$\kappa$
 plane for $m_t^{\rm pole} = 175$ GeV, $M_{\rm SUSY} = 350$ GeV, maximal stop
 mixing and the other parameters as in Eqs.~(\ref{param}, \ref{M2range}).}
\label{lkcontours}
\end{center}
\end{figure}

In Fig.~\ref{lkcontours}, we plot the maximal value of 
$\varepsilon_K$ in distinct regions of the $(\lambda, \kappa)$ plane.
The remaining parameters are chosen so to maximise $\varepsilon_K$ and
still comply with the experimental bounds. 
As one can see from this figure, having $\varepsilon_K \sim 2\times 10^{-3}$
is associated with values of $\kappa$ and $\lambda$ in the range
$[0, 0.6]$. In other words, one can easily saturate $\varepsilon_K$ in
a vast region of the singlet parameter space.
\begin{figure}
\begin{center}
\begin{tabular}{cc}
\includegraphics[height=42mm]{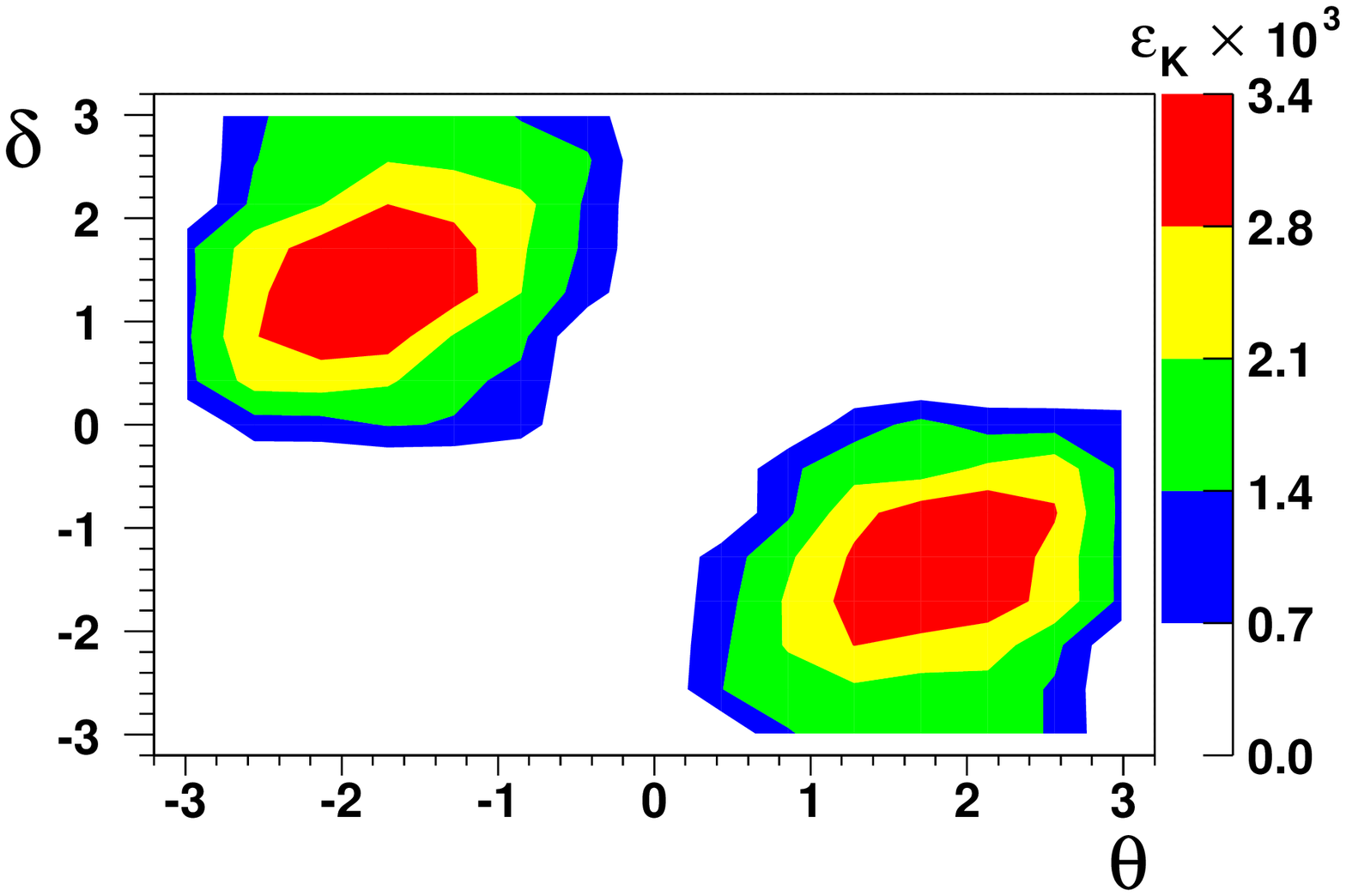} \hspace*{2mm}
&
\includegraphics[height=42mm]{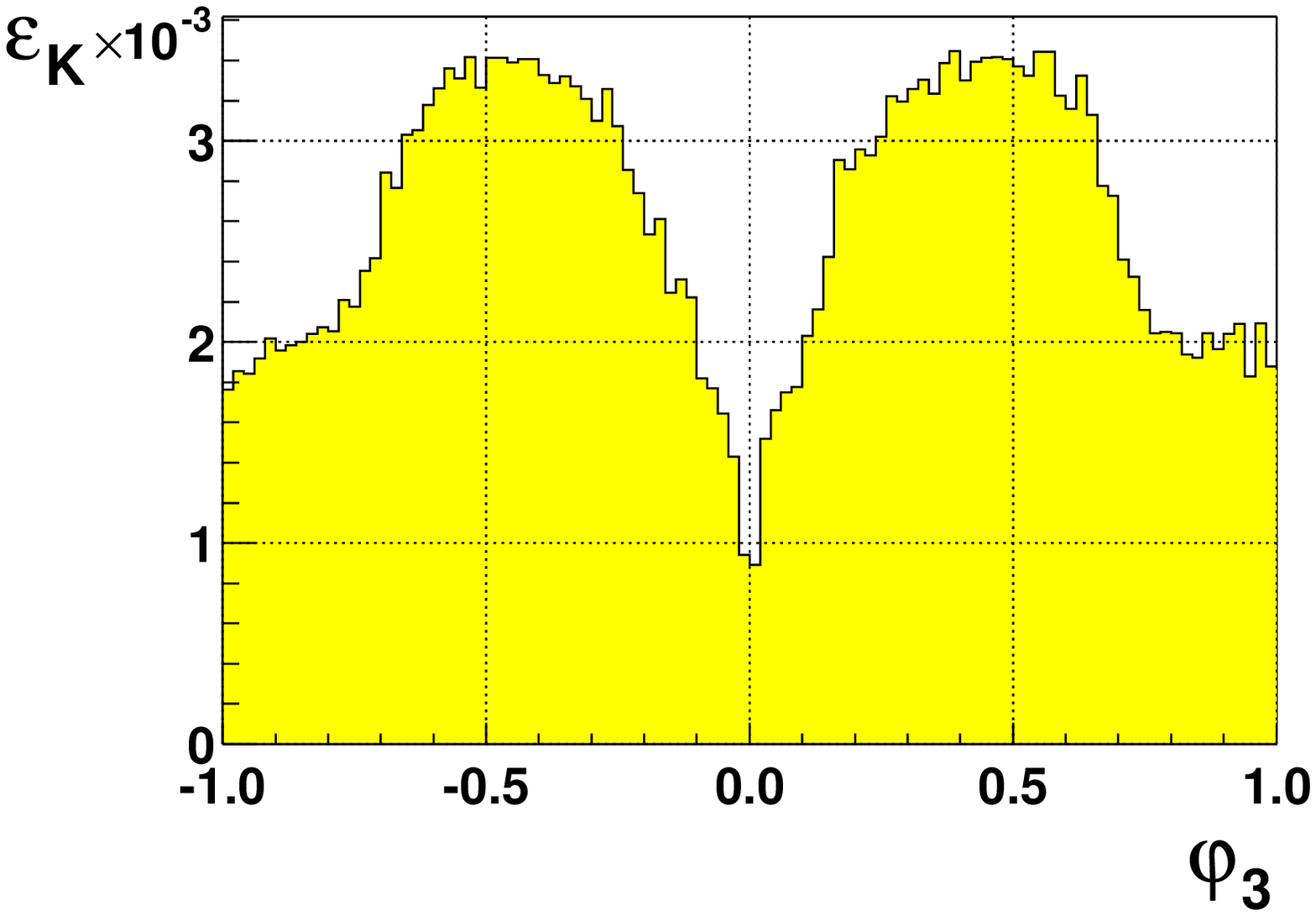}
\end{tabular}
\caption{Contours for the maximum value of $\varepsilon_K$ in the
  $\theta$--$\delta$ (left hand-side). 
Values of $\varepsilon_K$ as function 
of the singlet phase $\varphi_3=\delta/3$, depicted by the solid area
  (right hand-side). All parameters as in Fig.~\ref{lkcontours}.}
\label{PNeK}
\end{center}
\end{figure}
As expected from the inspection of Eq.~(\ref{eK:M12:equation}), there is a
strong dependence of $\varepsilon_K$ 
on the phases associated with the Higgs VEVs. In
Fig.~\ref{PNeK} (left hand-side), 
we display contour plots for the maximal values of 
$\varepsilon_K$ in the plane generated by the phases $\theta$ and
$\delta$. Although, a priori, all the values for the phases $\theta$ 
and $\delta$ in $[-\pi,\pi]$ are allowed, it is clear
from Fig.~\ref{PNeK} that the saturation of the experimental value of
$\varepsilon_K$ can only be achieved for significant values 
of the singlet and doublet phases. 
In Fig.~\ref{PNeK} (right hand-side) we show the values of $\varepsilon_K$ 
as a function of the singlet phase $\varphi_3=\delta/3$. 
The saturation of the experimental value of $\varepsilon_K$ 
requires the singlet phase to be $|\varphi_3| \gtrsim 0.15$. 

The consequences of such large CP violating phases are not negligible.
Let us recall that $\theta$ and $\delta$ are flavour-conserving
phases, and might
generate sizable contributions to the electron, neutron and mercury atom EDMs.
Although we will not address the EDM problem here, a few remarks are in order:
first, let us notice that in the presence of a small singlet coupling 
$\lambda$, as allowed in our results (see Fig.~\ref{lkcontours}), 
the EDM constraints on $\delta$
become less stringent~\cite{tanimoto}. In addition, there are several possible
ways to evade the EDM problem, namely reinforcing the non-universality on the
trilinear terms (\ie requiring the diagonal terms to be much smaller than the
off-diagonal ones or having matrix-factorisable $A$ terms), the existence of
cancellations between the several SUSY contributions, and the suppression of
the EDMs by a heavy SUSY spectrum~\cite{Khalil}. In view of the considerably
large parameter space allowed in our results, none of these possibilities
should be disregarded.

Finally, and concerning the other CP-violating observables, namely 
$\varepsilon^\prime/\varepsilon$ and the CP asymmetry of the 
$B_d$ meson decay ($a_{J/\psi K_S}$), it has been pointed
out~\cite{Pomarol2, Lebedev} that this class of models can generate
sizable contributions, although saturating the experimental values
generally favours a regime of large phases and maximal $LR$ squark
mixing. 

\section{Conclusions}\label{sect:5}
Spontaneous CP violation is a very appealing scenario, 
strongly motivated by both high- and low-energy arguments. Even though
several models have been considered, finding a consistent framework
for SCPV is not a trivial task.


The NMSSM with an extra tadpole term in the scalar potential appears
to be an excellent candidate for a SCPV scenario. Having a $\mathbb{Z}_3$
invariant superpotential preserves the original motivation of the
NMSSM to solve the $\mu$ problem of supersymmetry. The tadpole term
cures the domain wall problem, and allows the spontaneous
breaking of CP.
In this model, one can simultaneously saturate
$\varepsilon_K$ and obtain a sparticle spectrum compatible with
current experimental bounds. The analysis of the EDM is certainly
critical, and might prove to be a major viability test. 
 
Whether or not CP is explictly or spontaneously broken is a question
that must still be aswered. Even though the SM is a most accomplished 
effective theory, there are
strong reasons to believe that it is not the ultimate model of particle
physics. It is possible that nature has elected a realisation that
includes explicit CP violation, as is the case of the SM. Even so,
additional sources of CPV must arise from new physics. 
As we have tried to argue here, the hypothesis of spontaneous CP
violation is extremely appealing. Even though recent studies favour the
existence of a non-flat unitarity triangle (thus disfavouring
SCPV models), SCPV should not be ruled out. 
Ultimately, the advent of the LHC and a new generation of colliders
will be instrumental in addressing how CP is broken, and which model
of particle interactions (SM vs SUSY, minimal or non-minimal)
correctly describes nature.


\section*{Acknowledgements}

This work was supported by {\it Funda\c c\~ao para a Ci\^encia e
  Tecnologia} under the grant SFRH/BPD/11509/2002, and is based on
  collaborations with G. C. Branco, C. Hugonie, F. Kr\"uger and 
J. C. Rom\~ao. The author thanks the Organisers for the kind invitation to
  contribute to the Proceedings, and devotes a special acknowledgment
  to Ph.D. advisor and collaborator G. C. Branco, wishing him many
  more brilliant years pursuing an answer to the ``CP Violation and
  Flavour Puzzle''.

\end{document}